# Magnetic properties of the antiferromagnetic spin-½ chain system $\beta$-TeVO$_4$

Yu. Savina[1], O. Bludov[1], V. Pashchenko[1*], S.L. Gnatchenko[1], P. Lemmens[2], and H. Berger[3]

[1] *B.I. Verkin Institute for Low Temperature Physics and Engineering, NASU, 61103 Kharkov, Ukraine*
[2] *Institute for Condensed Matter Physics, TU Braunschweig, D-38106 Braunschweig, Germany*
[3] *Inst. Phys. Mat. Complexe, EPFL, CH-1015 Lausanne, Switzerland*

The magnetic susceptibility and magnetization of high quality single crystal $\beta$-TeVO$_4$ are reported. We show that this compound, made of weakly coupled infinite chains of VO$_5$ pyramids sharing corners, behaves as a $S = ½$ one-dimensional Heisenberg antiferromagnet. From magnetic experiments we deduce the intrachain antiferromagnetic coupling constant $J/k_B = 21.4\pm0.2$ K. Below 5 K a series of three phase transitions at 2.26, 3.28 and 4.65 K is observed.

**I. INTRODUCTION**

Vanadium oxides with the V$^{4+}$ ions ($S = ½$) are excellent model systems for one-dimensional spin-½ quantum magnets. Most studies have been focused on two classes of vanadium oxide chains with edge- and corner-shared VO$_n$ polyhedra (VO$_5$ square pyramids or VO$_6$ octahedra). Compounds in which adjacent building blocks share their edges are good realizations of the one-dimensional (1D) Heisenberg Hamiltonian.[1] In compounds built up of the corner-sharing topology, the nearest-neighbor (NN) exchange coupling may be more then an order of magnitude smaller[2,3] and an important role of the next-nearest neighbor (NNN) interactions take place. The interplay between frustration and quantum spin fluctuations results in a rich phase diagram with commensurate/incommensurate phases and unusual magnetic properties of these materials.

Recently, $\alpha$-TeVO$_4$ with the zigzag chains formed by distorted VO$_6$ octahedra sharing edges have been discovered as a quasi-one-dimensional spin-½ Heisenberg system with alternating NN ferromagnetic (FM) interactions and NNN antiferromagnetic (AF) interaction.[4] As it was known[5], the compound TeVO$_4$ can be prepared in two different crystalline forms with a reversible polymorphic transformation at 650°C ($\alpha\leftrightarrow\beta$). In contrast to $\alpha$-phase, $\beta$-TeVO$_4$ has a significant difference in its structure: the zigzag chains are formed by distorted VO$_5$ pyramids sharing their corners. Thus, the corner-sharing of VO$_5$ pyramids can lead to the considerable reducing of a scale of the existing exchange interactions in the 1D spin system.

In this paper, we report on the single crystal magnetic properties of a quasi-one-dimensional spin-½ chain system, $\beta$-TeVO$_4$. Though this compound phase has been known for a long time[5], to our knowledge, its magnetic properties on powdered or single crystal samples were hitherto not reported at all. We discuss that the actual topology of magnetic couplings for such system might be indeed more complex than originally envisioned.

**II. EXPERIMENTAL DETAILS**

The growth of $\beta$-TeVO$_4$ single crystals was described elsewhere[5]. Sample with dimensions of 3×2×4 mm$^3$ was oriented by using a X-ray Laue diffractometry. Variable-temperature magnetic susceptibility measurements in the temperature range 1.9 – 400 K and magnetic fields 0.001 – 0.1 T along three crystallographic axes were carried out on a single crystal sample of $\beta$-TeVO$_4$ ($m$ = 128.33 mg) using a Quantum Design SQUID magnetometer MPMS-XL5. Susceptibility measurements were complemented by isothermal magnetization runs at temperatures between 2 and 50 K for fields up to 5 T.

**III. RESULTS AND DISCUSSION**
**A. Structure description and important parameters**

The compound $\beta$-TeVO$_4$ crystallizes in the monoclinic system with the space group P2$_1$/c and the parameters: $a$ =4.379±0.002 Å, $b$ =13.502±0.004 Å, $c$ =5.446±0.002 Å and $\beta$ =91.72°±0.05° with Z=4 (f.u. TeVO$_4$ per unit cell).[5] The crystal structure of $\beta$-TeVO$_4$ is shown in Fig. 1. The structure consists of zigzag chains (double chains) parallel to the $c$ axis formed by slightly distorted square pyramids VO$_5$ sharing corners. There are two identical zigzag chains, in which all apices of square pyramids are pointing alternatively below and above the *bc* plane. The lone pair cation Te$^{4+}$ leads to a magnetic separation of chains with respect to each other. A twofold axis $C_2$ is directed along the *b* axis of crystal. In accordance to the symmetry properties there is only one symmetry-independent position for vanadium atom and only one NN exchange path for the V$^{4+}$ ions through the corner oxygens (the V–O–V angle is 133.7°) into a chain is expected. The nearest intrachain distance is V–V = 3.6427 Å. The nearest V–V distances perpendicular to the chain direction are 4.9149 (along the *b* axis) and 4.3790 Å (along the *a* axis), respectively.

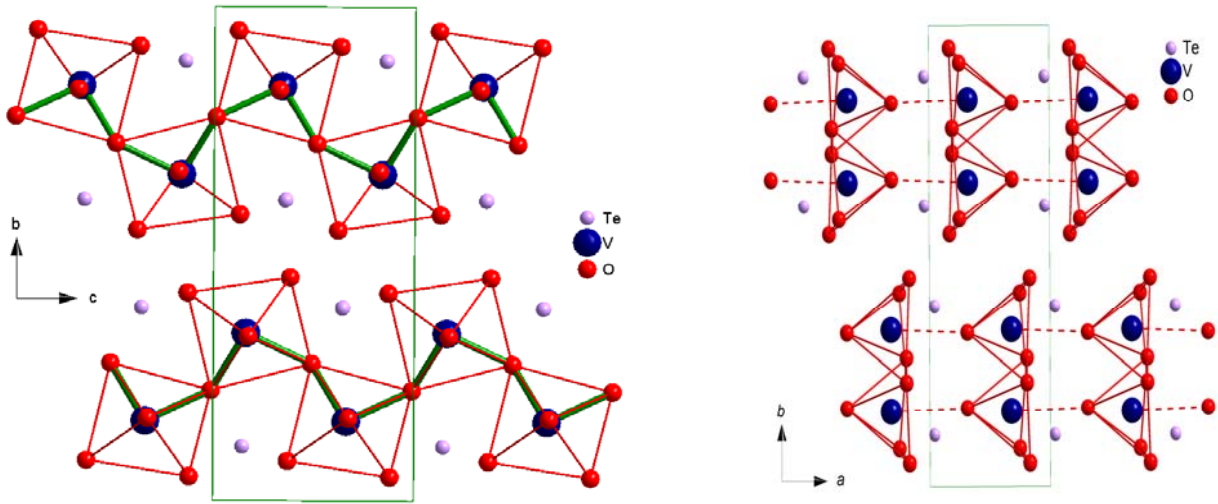

FIG. 1. (Color online) Crystal structure of *β*-TeVO$_4$, viewed along *a* (left) and *c* (right) axis. The corner-sharing paths in a zigzag chain responsible for nearest-neighbor exchange interactions are shown by the thick green lines. The V–O bonds between the nearest VO$_5$ pyramids along the *a* axis are shown as dashed lines.

Figure 2 shows the nearest oxygen environment of a vanadium atom in the crystal structure *β*-TeVO$_4$. Each vanadium atom is surrounded by five nearest-neighbor oxygen atoms forming a slightly distorted square pyramid VO$_5$. A sixth oxygen neighbor O$_{ap}$´, which can be also interpreted as an apical oxygen of the next VO$_5$ pyramid, is found at a much longer distance (V–O$_{ap}$´ = 2.772 Å, ∠(O$_{ap}$–V–O$_{ap}$´) = 173.6°) in the position usually issued from a highly distorted octahedron VO$_6$ (see Fig. 1, 2). The apical distance V–O$_{ap}$ (1.613 Å) is much smaller than the V–O distances involving either the oxygens contained in the pyramid basal plane (1.927-2.035 Å), or the sixth oxygen O$_{ap}$´ (2.772 Å). The presence of additional long bond V–O$_{ap}$´ might give a possibility to ascribe the nearest environment of the vanadium atoms as a highly distorted octahedron, but the obtained large shift of 0.515 Å for a vanadium atom from the basal oxygen plane is in favor to a square pyramidal environment. Additionally note that a typical range of the longest apical bonds for a distorted octahedron VO$_6$ is usually $2.10 < d(\text{V}^{4+}–\text{O}_{ap}) < 2.70$ Å.[6] In the case of the "octahedron presentation", the crystal structure of *β*-TeVO$_4$ might be regarded as a two-dimensional framework with well separated layers of the distorted VO$_6$ octahedra perpendicular to the *b* axis. As it was shown in Ref. 7, the exchange pathway through common apical corners of VO$_6$ octahedra is leading to a very weak magnetic interaction between the V$^{4+}$ ions with a magnitude smaller then 2 K and therefore the pronounced two-dimensional magnetic behavior of *β*-TeVO$_4$ is not expected.

Each square pyramid VO$_5$ may actually be seen as a vanadyl V≡O$_{ap}$ cation lying above a slightly distorted square of O$^{2-}$ anions. The apical V–O$_{ap}$ bond is not rigorously perpendicular to the basal oxygen plane of the square pyramids (the angle is 87.831°). As it was mentioned in Ref. 8, the existence of an apical oxygen O$_{ap}$ is crucial in such compounds since the shortest V–O$_{ap}$ distance allows the possibility of a strong delocalization to occur between these two atoms and a multiple vanadium-oxygen covalent bond to take place. The experimental signature of such a strong multiple bond in the Raman spectra is the presence of a sharp peak at relatively high energy, corresponding to the stretching mode. This peak, for example, occurs at 969 cm$^{-1}$ for the α´-NaV$_2$O$_5$ [9], at 932 cm$^{-1}$ and 1002 cm$^{-1}$ for the CaV$_2$O$_5$ and MgV$_2$O$_5$ [10], respectively.

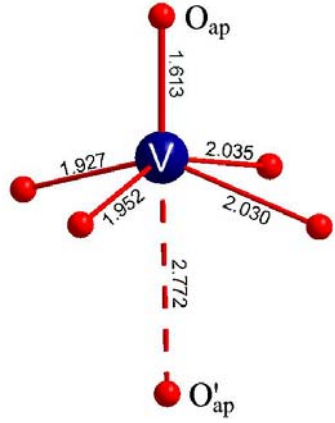

FIG. 2. (Color online) Oxygen environment of the V$^{4+}$ ion in crystal structure of β-TeVO$_4$. The length of V–O bonds is shown in Å.

The corner linkages of the VO$_5$ pyramids in the chain of β-TeVO$_4$ revealed here are almost identical to those of the zigzag chain compounds CdVO$_3$ [2] and CaV$_2$O$_5$ [3] formed by sharing edges and corners of VO$_5$ pyramids. The V–O–V angles for the corner-sharing paths in CdVO$_3$ and CaV$_2$O$_5$ are 136.1° and 135.3°, respectively. Both compounds have a nearest-neighbor V–V distance is around of 3.60 Å. The magnetic study of the zigzag chain of CdVO$_3$ has demonstrated that the corner-sharing exchange coupling is a quit small and is not relevant to magnetic properties of this compound. This statement is also consistent with an isolated dimer model for the zigzag chain compound CaV$_2$O$_5$. Therefore, the NN exchange interactions (see $J_1$ in Fig. 3), which are due to corner-sharing paths in the chain of β-TeVO$_4$, are expected to be weak in a magnitude. In this case we should take also into account the possible NNN exchange paths like V–O–O–V (see $J_2$ in Fig. 3) with connectivity through two oxygens of the basal plane. Thus, an effective 1D magnetic model might be looked also as a network of isolated two-leg ladders (double chains) with two different exchange integrals $J_1$ and $J_2$ ($J_1$-$J_2$ model). The period of each "leg" of the double spin chains is equal to $c$. The two legs are offset by the $c/2$ relative to each other and thus form a "triangular ladder". In purely 1D case, depending on the sign and magnitudes of $J_1$, $J_2$, the ground state of the isolated $J_1$-$J_2$ model might be either the quantum disordered gapped state (with commensurate or incommensurate spin correlations) or the gapless state with commensurate spin correlations. Note, for $J_1 \gg J_2 \to 0$ or $J_2 \gg J_1 \to 0$, the $J_1$-$J_2$ model transforms to a uniform spin-½ chain system. In case of weakly coupled chains in three dimensions, a large variety of magnetic ground states of the system might be realized: a disordered gapped state (nonmagnetic at $T = 0$), a collinear long-range Neel state, a helimagnetic long-range order and, finally, in a certain temperature range, a chiral long-range order with broken symmetry of left/right spirals and helimagnetic incommensurate short-range correlations without magnetic long-range order.

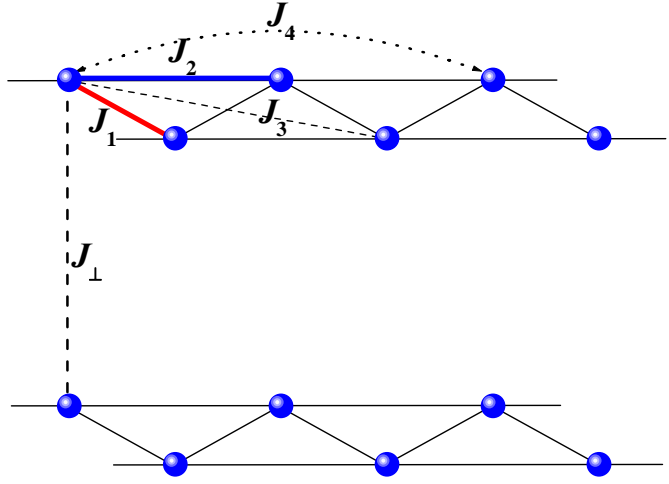

FIG. 3. (Color online) A schematic view of exchange interactions between the $V^{4+}$ ions in $\beta$-TeVO$_4$. The most important magnetic interactions are sketched.

Prolonging the further model complications, similarly to the "NNN" case we can even consider the nearest-neighbor pathways of the higher order: *third-* or *fourth*-nearest-neighbor exchange interactions through the V–O–O–O–V or V–O–O–O–O–V bridges, respectively. A total scheme of possible exchange interactions between the $V^{4+}$ ions in $\beta$-TeVO$_4$ is illustrated in Fig. 3. It should be noted that for a $J_1$-$J_2$ model the presence of the third interchain interaction $J_\perp$ (possibly quite weak in its amplitude) through the bridge V–O$_{ap}'$–V between neighboring pyramids in $\beta$-TeVO$_4$ and/or the existence of the nonzero *third-* or *fourth*-nearest-neighbor coupling constant (see $J_3$ or $J_4$ in Fig. 3) gives a total similarity with the exchange model topology, which are proposed and discussed for quasi-one-dimensional helimagnets, LiCu$_2$O$_2$ [11] and NaCu$_2$O$_2$ [12].

Thus, the structural data allow us to consider the compound $\beta$-TeVO$_4$ as a uniform spin-½ chain system (for the small exchange magnitudes possibly with competing NN and NNN interactions, an alternating AF/FM exchange interactions, anisotropic and antisymmetric Dzyaloshinskii-Moriya interactions).

**B. Magnetic susceptibility and magnetization**

The magnetic susceptibility $\chi(T) = M(T)/H$ of $\beta$-TeVO$_4$ single crystal ($M$ denotes the magnetic moment of the sample) measured along three principal crystallographic $a$, $b$, $c$ axes are shown in Fig. 4. The inverse magnetic susceptibility $\chi^{-1}(T)$ and the product $\chi(T)T$ for the same data are shown in the insert of Fig. 4 and in Fig. 5, respectively. The magnetic data were found to be reversible upon cooling and heating with no indications for hysteretic behavior. The collapse of the susceptibility data onto a single curve for various measured fields indicates that no significant field dependence occurs over the whole temperature (5–400 K) and field (0.001–0.1 T) range under investigation. We have estimated a diamagnetic contribution from the closed atomic shells for $\beta$-TeVO$_4$ as $-9.5 \cdot 10^{-5}$ cm$^3$ mol$^{-1}$ by using tabulated values for Pascal's constants.[13]

In Fig. 4 magnetic susceptibility data shows a noticeable axial anisotropy of magnetic properties of single crystal $\beta$-TeVO$_4$ respect to the crystallographic $b$ axis. The ratio between magnetic susceptibility for $H \| b$ and $H \perp b$ keeps a constant value of $\chi_\| / \chi_\perp$ ($\chi_\| > \chi_\perp$) in the whole temperature range. For magnetic fields applied within $ac$ plane ($H \perp b$) we did not find any significant difference in magnetic behavior of the studied system. The most probable explanation of such magnetic anisotropy is due to an axial symmetry of the $g$-tensor for the $V^{4+}$ ions with a square-pyramidal environment. For example, the estimated $g$-values extracted from Curie constants (see in Table I) give a quite well agreement with the expected ones from the literature.[14] The fact that the $b$ axis is a preferential magnetic direction of the crystal might be contradicted with the obtained crystallographic data, especially with the spatial orientation of the VO$_5$ pyramids, in which all apical axes (a short V–O$_{ap}$ bond) are located along the crystallographic $a$ axis. To clarify this question it is necessary additionally to perform the electronic structure calculations in order to identify the real spatial orientation of the vanadium 3$d$ magnetic orbitals in the crystal structure of $\beta$-TeVO$_4$.

The main feature of three $\chi(T)$ curves for $a, b, c$ axes is the presence of a maximum at $T\chi_{max}$ =14.0±0.1 K (a temperature at which $d\chi/dT = 0$), which is characteristic for low-dimensional spin systems. Note, the magnetic susceptibility of an uniform antiferromagnetic Heisenberg spin-½ chain is characterized by a maximum of $\chi$ at a temperature $T\chi_{max} = 0.640851 \cdot |J_{1D}|/k_B$ with a value $\chi_{max} = 0.146926 \cdot Ng^2\mu_B^2/|J_{1D}|$.[15, 16] By using these relations we can easily estimate an expected AF exchange coupling as $|J_{1D}|/k_B$ = 21.8 K and an axial g-tensor with g = (2.188; 2.285; 2.196), which is a little bit larger then expected for the V$^{4+}$ ions. From high temperature series expansion of $\chi(T)$, the negative AF Curie-Weiss temperature $\theta_{1D}$ = -½$|J_{1D}|/k_B \approx$ -10.9 K in the Curie-Weiss law $\chi(T) = C/(T - \theta_{1D})$ is expected.

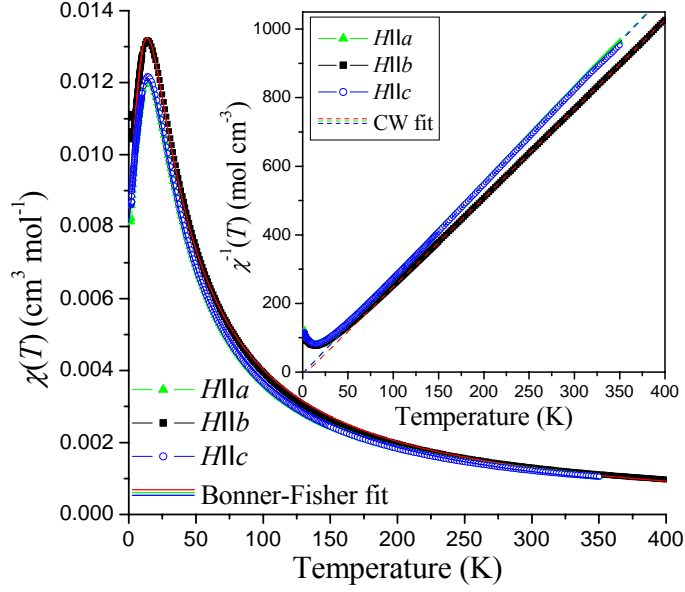

FIG. 4. (Color online) Temperature dependence of molar magnetic susceptibility $\chi(T)$ of single crystal $\beta$-TeVO$_4$ measured in a magnetic field $H$ = 0.02 T applied along three crystallographic axes. The solid lines show the best fit for an uniform AF spin-½ chain (Bonner-Fisher model) with parameters specified to Table II. The insert shows temperature dependence of the inverse magnetic susceptibility $\chi^{-1}(T)$ and the best fit by using a Curie-Weiss law (dashed lines, examined in the range 150–400 K).

The additional details of the high temperature behavior become more visible in a plot of $\chi(T)T$ vs $T$ presented in Fig. 5. A closer look at the product $\chi(T)T$ gives an indication of the presence of the weak ferromagnetic spin-spin correlations down to $T_{Cros}$ =130 K: the value of $\chi(T)T$ continuously increases on cooling to reach a broad maximum at $T_{Cros}$ =130 K and then rapidly decreases when the temperature is lowered, which is the typical nature of significant antiferromagnetic exchange interactions between the magnetic centers. A broad crossover at $T_{Cros}$ separates two different magnetic regime with predominant ferro- and antiferromagnetic correlations above and below this temperature. Note, such scenario is possible only for the complex topology of the exchange pathways in low-dimensional spin systems. In addition, there are two features at 315 K (***H***⊥*b*) and at 380 K (***H***∥*b*) marked by vertical arrows in the insert of Fig. 5, which might be due to other magnetic instabilities and requires a further investigation.

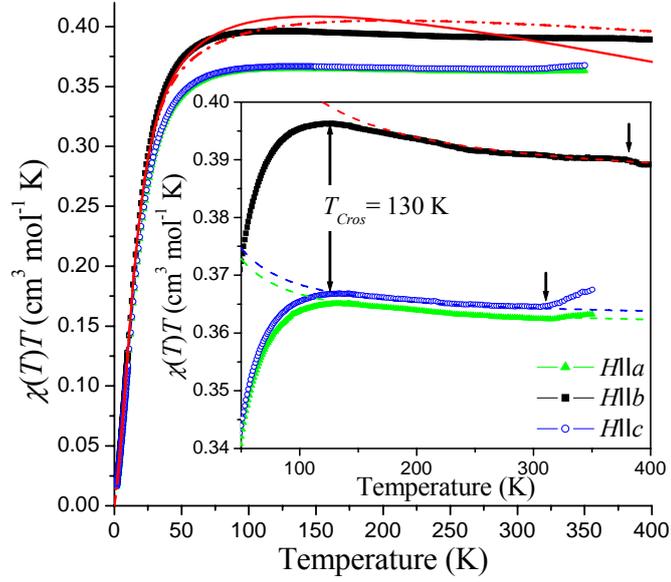

FIG. 5. (Color online) The product $\chi(T)T$ of single crystal $\beta$-TeVO$_4$ as a function of temperature $T$ for three crystallographic directions. The solid line shows the result of a fit to the uniform 1D chain model, whereas the dot-dashed line is the best fit for the weakly-coupled AF chain model ($\boldsymbol{H} \parallel b$). The insert shows the enlarged scale of high temperature behavior. The dashed lines are the results for Curie-Weiss law with magnetic parameters specified to the Table I.

*1. Magnetization*

Magnetization curves measured in the temperature range 2–50 K in the applied fields along three crystallographic axes were well proportional to the applied magnetic field, which inferred that a significant spin gap is not present in this system. No any remanent magnetization at $H=0$ is detected. Figure 6 shows the magnetization $M(H)$ vs $H$ along the $b$ axis at varying temperatures from 2–50 K.

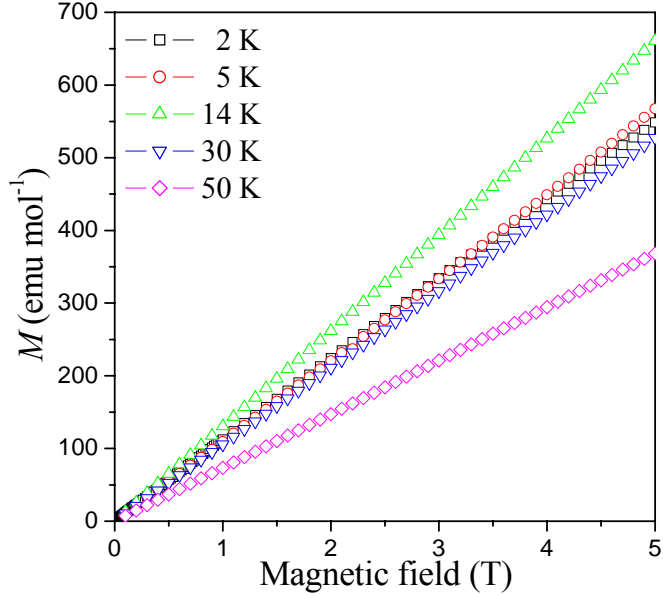

FIG. 6. Isothermal magnetization measurements in dc fields from 0–5 T along the $b$ axis at varying temperatures as indicated in the figure.

From isothermal magnetization measurements in magnetic fields 0–5 T at varying temperatures, we have extracted the values for the differential magnetic susceptibility $\left(\dfrac{dM(T)}{dH}\right)_{H \to 0}$, which can be compared with magnetic data presented in Fig. 4. The collapse of the susceptibility data is absolute. Figure 6 shows a perfect linear regime of the magnetization $M(H)$ at least above 5 K.

*2. Low temperature behavior of magnetic susceptibility (below 5K)*

Low temperature behavior of magnetic susceptibility $\chi(T)$ of single crystal $\beta$-TeVO$_4$ measured in a magnetic field along *a*, *b*, *c* axes are presented in Fig. 7. This is an enlarged low temperature part of data shown in main panel of Fig. 4. Below 5 K the presence of three different magnetic features at $T = 2.26$; 3.28; 4.65 K are observed. The first two at 3.28 and 4.65 K look like a kink on $\chi(T)$. The high temperature kink is better visible for $\mathbf{H}\perp b$, while the low temperature kink is more pronounced for $\mathbf{H}\| b$. A closer look at the derivative $d\chi/dT$ vs $T$ (see insert in Fig. 7) gives us a proof that the both these features simultaneously exist for any orientation of measured field. The third feature at 2.26 K looks as a jump (with a step-like shape) on $\chi(T)$ with much stronger amplitude then the previous two. Note, there is a "sign" changing of the susceptibility jump from "–" to "+" for directions perpendicular and along the *b* axis. All three features do not show any significant hysteresis behavior upon cooling and warming up (see, for instance, the curve for $\mathbf{H}\| c$). All three features might be interpreted as a series of magnetic phase transitions in a low dimensional spin system. But the nature of each phase transformation is unknown. Taking into account the predominant character of antiferromagnetic correlations below 130 K, we can assume that the compound at $T_N = 4.65$ K undergoes a magnetic phase transition into a long-range-ordered AF magnetic state and the next two features are due to the further modifications of antiferromagnetic ordered phase. Such scenario containing a series of few phase transitions pointing to a complex multi-stage rearrangement of the spin structure was reported for quasi-one-dimensional helimagnetic systems, LiCu$_2$O$_2$ [17] and NaCu$_2$O$_2$ [12]. Specific heat measurements are better suited to unambiguously identify $T_N$ and the nature of other two features in $\beta$-TeVO$_4$ and are in progress.

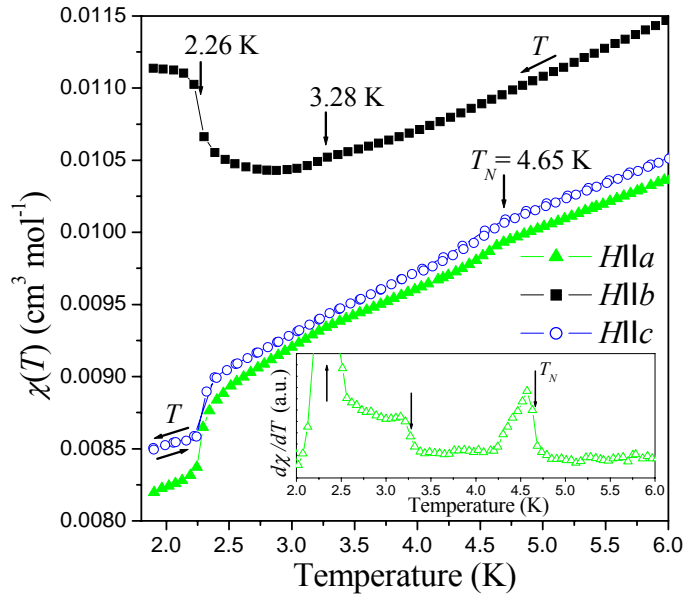

FIG. 7. Low temperature behavior of magnetic susceptibility $\chi(T)$ of single crystal $\beta$-TeVO$_4$ measured in a magnetic field $H = 0.02$ T applied along three crystallographic axes. The vertical arrows indicate three observed magnetic features. The insert shows $d\chi/dT$ vs $T$ for $\mathbf{H}\| a$.

Note, the absence of any characteristic upturns of $\chi(T)$ in low temperature range (below 2.2 K) indicates that the Curie tail is very weak in $\beta$-TeVO$_4$ and is providing the testimony to the high quality of the single crystal without any impurity or defects.

*3. Analysis*

The analysis of the magnetic properties of $\beta$-TeVO$_4$ has been performed in the following three steps. First, in the high-temperature range 150–400 K, we have used a Curie-Weiss law to obtain preliminary information on the sign and average magnitude of the exchange correlations between the V$^{4+}$ ($S=\frac{1}{2}$) ions. Then, we have made an attempt to analyze the obtained experimental data within the framework of few microscopic models with an uniform exchange coupling constant (FM or AF) between nearest-neighbor spins into the chain. In a third step, we have considered several 1D models, which are "exotic" for given case. The main purpose of this analysis was to find a magnetic model yielding a good qualitative and quantitative agreement with experimental data in the wide temperature range (above the long-range ordering $T_N$).

*(a) High temperature range.* In a first attempt at setting up a simple model, capable of describing the main features of the magnetic susceptibility of $\beta$-TeVO$_4$ at high temperatures above 130 K, magnetic behavior can be assigned to Curie-Weiss-like temperature dependence $\chi(T) = C/(T - \theta)$ with a Curie constant $C$ and a Curie-Weiss temperature $\theta$. A Curie-Weiss fit of magnetic susceptibility in the temperature range 150 K<$T$<400 K yields a Curie constant $C_b$ = 0.3852 cm$^3$ mol$^{-1}$ K and a positive Curie-Weiss temperature of $\theta_b$ = +4.418 K for magnetic field applied along the $b$ axis. For magnetic fields applied within the $ac$ plane ($\mathbf{H}\perp b$), we find $C_a$ = 0.3609 cm$^3$ mol$^{-1}$ K and $\theta_a$ = +1.598 K (along the $a$ axis) and $C_c$ = 0.3623 cm$^3$ mol$^{-1}$ K and $\theta_c$ = +1.658 K (along the $c$ axis), respectively. The results are summarized in Table I. The obtained Curie constant is consistent with the presence of one V$^{4+}$ ion ($S=\frac{1}{2}$) per formula unit yielding a spin-only value of $C_{theory}$ = 0.3751 cm$^3$ mol$^{-1}$ K. The latter number is based on a spectroscopic $g$-value of 2.0. It should be noted that the $g$-tensor extracted from the obtained Curie constants ($C = \dfrac{N\mu_B^2 g^2 S(S+1)}{3k_B}$, where $N$ is Avogadro's number, $\mu_B$ is the Borh magneton, $k_B$ is the Boltzmann constant, $S = \frac{1}{2}$ and $g$-value for the V$^{4+}$ ions) has an axial anisotropy, and its components ($g_\perp$, $g_\parallel$) are in a satisfactory agreement to expected ones from the literature (see Table I). The expected range of the spectroscopic $g$-values for the V$^{4+}$ ions within a square-pyramidal oxygen environment is 1.94–1.98.[14] Note, such anisotropy suppose that the magnetic orbitals of the V$^{4+}$ ions with a square-pyramidal environment are orthogonal to a very short vanadyl bond V–O$_{ap}$ (and to a zigzag chain direction too). The positive Curie-Weiss temperatures indicate possible predominant ferromagnetic character of interactions with quit small amplitudes (or a sum of AF and FM interactions with the almost similar exchange strength). It is possible to discuss a small axial anisotropy of these effective ferromagnetic interactions, which follows from the different Curie-Weiss temperatures of the system along and perpendicular to the $b$ axis.

Table I. Effective magnetic parameters for a Curie-Weiss model used in high temperature range (150-400K).

| Field direction | Curie constant $C$ (cm$^3$ mol$^{-1}$ K) | Curie-Weiss temperature $\theta$ (K) | $g$-value for the V$^{4+}$ ion extracted from $C$ |
|---|---|---|---|
| $H\parallel b$ | 0.3852 | +4.418 | 2.027 |
| $H\parallel a$ ($H\perp b$) | 0.3609 | +1.598 | 1.962 |
| $H\parallel c$ ($H\perp b$) | 0.3623 | +1.658 | 1.965 |

At a first glance, the fit by using a simple Curie-Weiss law agrees quite well with the experimental data in the high-temperature region (see the dashed lines in insert of Fig. 4, 5). However, it should be noted that a simple Curie-Weiss law can not explain the strong decrease in $\chi(T)T$ below $T_{Cros}$ = 130 K, which is clearly indicative to the dominant antiferromagnetic interactions in this compound. Thus, for low temperature range the search of an appropriate microscopic magnetic model would be desirable.

*(b) Uniform exchange coupling model.* The uniform Heisenberg spin-½ chain with a FM or AF coupling is a magnetic model, which is frequently realized in low dimensional solids. The isotropic AF spin-½ Heisenberg chain was first diagonalized by Bethe.[18] The spin Hamiltonian of the uniform Heisenberg spin-½ chain $H = -J\sum_i \vec{S}_i \cdot \vec{S}_{i+1}$ (for $J > 0$ and $J < 0$) was examined by Bonner and Fisher (1964).[16] Following these works many important properties with the highest accuracy for this model have been obtained by several authors.[15, 19]

Taking into account only the structural data and without additional knowledge about the studied system, it is difficult to predict the exact relationship and the signs of all exchange integrals {$J_1$, $J_2$, $J_3$, $J_4$} in the zigzag chain of $\beta$-TeVO$_4$. Let's discuss a $J_1$-$J_2$ model ($J_1$, $J_2 \gg J_3$, $J_4$), where $J_1$ and $J_2$, for instance, are the NN (V–O–V) and NNN (V–O–O–V) exchange paths (see Fig. 3). The presence of a broad crossover may provide a different sign for two these interactions. Under the conditions $|J_1| \gg |J_2|$ or $|J_1| \ll |J_2|$, the zigzag chain has the main properties of a uniform FM or AF spin-½ Heisenberg chain with the smallest exchange as a model extension. It should be noted that the model of an uniform FM spin chains coupled by a weak AF interaction (FM $|J_1| \gg$ AF $|J_2|$) can not be applied for our case because for this model the maximum of ferromagnetic susceptibility $\chi_{max}$ and its temperature position $T\chi_{max}$ may have a strong field dependence as expected.[16] In contrast to this, our experimental data doesn't show any field dependence on the magnetic susceptibility of $\beta$-TeVO$_4$. Thus, we will try to assume that the sign of dominant exchange is presumably AF, while the next one is probably FM (FM $|J_1| \ll$ AF $|J_2|$). This assumption, as well as taking into account the possible influence of other weak interactions $J_3$, $J_4$ and $J_\perp$, suggests that the magnetic interactions are frustrated and magnetic model is close to a system of weakly coupled antiferromagnetic chains ($|J_2| \gg |J_1|, |J_3|, |J_4|, |J_\perp|$). Thus, we will try to estimate two effective magnetic parameters: the dominant intrachain exchange coupling $J$ ($J \equiv J_2$) and the interchain interaction $J'$, which is the result of some combination of all weak interactions {$J_1$, $J_3$, $J_4$, $J_\perp$} and leads to the long-range magnetic order at low temperatures.

To determine the dominant antiferromagnetic exchange parameter $J$, we fit our magnetic susceptibility measurements with an expression of the form:

$$\chi(T) = \chi_{1D}(T) + \chi_{TIP}. \qquad (1)$$

$\chi_{1D}(T)$ is the contribution of the spin $S = ½$ Heisenberg antiferromagnetic chains, which is known with a high precision over the whole measured temperature range.[15] We took the polynomial approximation of Bonner and Fisher (a well-known Bonner-Fisher model):

$$\chi_{1D}(T) = \frac{Ng^2\mu_B^2}{k_B T} \frac{0.25 + 0.074975x + 0.075236x^2}{1 + 0.9931x + 0.172135x^2 + 0.757825x^3}, \qquad (2)$$

where $x = |J|/k_B T$. The presence of weak ferromagnetic correlations at the high temperatures was accounted for by using an additional temperature-independent parameter $\chi_{TIP}$ with a small negative value. The chosen value of $\chi_{TIP}$ is comparable in amplitude with the estimated value of diamagnetic contribution for $\beta$-TeVO$_4$. One can note that the proposed AF uniform exchange coupling model fits well the measured curves in the wide temperature range 5–400 K (see solid lines, mostly hidden by the experimental data, in Fig. 4). The best least squares fit parameters are summarized in Table II. The most important conclusions, which emerge from this analysis, are the following: i) the compound $\beta$-TeVO$_4$ exhibits a rather pronounced $S = ½$ quasi-1D antiferromagnetic behavior in the wide temperature range; ii) the value of the single antiferromagnetic *intrachain* exchange constant can be evaluated as $J/k_B = 21.4 \pm 0.2$ K, quite independently of the choice of principal crystallographic directions. There is only one open question: the obtained $g$-values (2.18-2.28) are much bigger then expected ones (1.94-1.98) from the literature and the estimated from high temperature analysis.

Table II. Magnetic parameters extracted by using Bonner-Fisher model in the temperature range 5–400K.

| Field direction | $J/k_B$ (K) | $g$ | $\chi_{TIP}$ ($\times 10^{-4}$ cm$^3$ mol$^{-1}$) |
|---|---|---|---|
| $H \parallel b$ | 21.20 | 2.280 | -2.5 |
| $H \parallel a$ ($H \perp b$) | 21.41 | 2.185 | -2.5 |
| $H \parallel c$ ($H \perp b$) | 21.56 | 2.200 | -2.5 |

On the next step of analysis we can introduce the further conjunction between the AF spin chains, which is due to the effective interchain interaction $J'$ ($J' \ll J$). The magnetic behavior of weakly-coupled chains in quasi-one-dimensional spin system can be described by a chain mean-field theory developed by Schulz.[20] This theory gives the relations between the ordered magnetic moment

$m_0$, the ordering temperature $T_N$, the main antiferromagnetic exchange coupling $J$ and the average of interchain interactions $|J'|$ in quasi-one-dimensional spin system. One of these relations allows an estimate for $|J'|$:

$$|J'| = \frac{T_N}{4 \cdot 0.32(\ln(5.8 J / T_N))^{1/2}}. \qquad (3)$$

In our case, where $T_N = 4.65$ K and $J/k_B = 21.4$ K, this gives $|J'|/k_B = 2.00$ K. The real value for $|J'|$ may be even somewhat higher because quantum fluctuations are not fully accounted for in the chain mean-field theory. Thus the ratio between interchain and intrachain coupling is $9.3 \times 10^{-2}$, showing that this compound is a quasi-one-dimensional spin-system.

In order to estimate the average amplitude and the sign of the possible interchain interaction $J'$ we will try to reanalyze the experimental data in the temperature range 5–400 K by using the following mean-field approximation for magnetic susceptibility:

$$\chi_{MF}(T) = \frac{\chi_{1D}}{1 + (2zJ'/Ng^2\mu_B^2)\chi_{1D}}, \qquad (4)$$

where $z$ is the number of the nearest neighbors. The important conclusions from this analysis are: i) the sign of the average *interchain* interaction $J'$ is negative, that is in favor of the weak ferromagnetic interchain correlations; ii) the best fit gives the parameter $z|J'|/k_B$ of ferromagnetic coupling of order of 0.3–0.9 K; iii) the presence of the weak *interchain* interactions is not affect on the value of the main antiferromagnetic *intrachain* coupling ($J \gg J'$) and does not bring a significant improvement of the description accuracy of experimental data in whole. Note, as it is demonstrated in Fig. 5, the description by using an isolated and weak-coupled chain models has a quite well agreement with experimental data at low temperatures and is not perfect above $T_{Cros} = 130$ K (see, for example, the solid and dot-dashed lines as the best fits for $H \| b$).

As one of the last possible extension of the uniform Heisenberg spin-½ chain model we can try to introduce an alternation of the AF interactions into a chain, which leads to the formation of a spin gap at low temperatures. The accurate theory for a spin-½ AF alternating-exchange Heisenberg chains was developed by D.C. Jonhston and coauthors.[15] The best fits with the alternation parameter $\alpha \geq 0.99$ gave the similar good accuracy of the experimental description like two previous cases. From this analysis we can conclude that we are not able to detect experimentally the presence of a spin gap of the 1D system if the exchange alternating parameter $\alpha \geq 0.92$ (for given AF exchange of 21.4 K). It should be noted that even below the ordering temperature $T_N = 4.65$ K the deviations of the experimental susceptibility data from the theoretical predictions for an uniform Heisenberg spin-½ chain are quite small and keep the tendency of the temperature evolution of a non-ordered phase.

*(c) Non-uniform exchange coupling models.* In a third step, we have considered several additional 1D models, which might be realized in the studied system with the small amplitudes of exchange interactions: Alternating Chain with FM and AF Interactions [21] and Bleaney-Bowers AF dimer [22]. Note, all these models yields no satisfactory agreement with the observed experimental data especially around of a maximum.

To summarize, we have found that an AF uniform exchange coupling model for spin $S=½$ chain (see solid lines in Fig. 4), even without the possible model extensions such as a weakly-coupled AF chains or an alternating-exchange AF spin-½ chain, is capable to reproduce satisfactory the magnetic behavior of the studied compound. The fact that only *single* parameter such as an *intrachain* AF coupling can well cover the main properties of the low-dimensional system with the complex topology of the exchange interactions is not surprising. As it was demonstrated for a quasi-one-dimentional helimagnet NaCu$_2$O$_2$ [12], the magnetic behavior of the helix model with four exchange parameters $\{J_1, J_2, J_3, J_4\}$ coincides surprisingly good to the susceptibility of the 1D AF Heisenberg chain under condition that the single AF coupling is dominant.

### IV. CONCLUSIONS

To summarize, we have presented magnetic susceptibility and magnetization of a quasi-one-dimensional spin-½ chain system, $\beta$-TeVO$_4$. Magnetic data for three crystallographic axes shows an axial anisotropy of magnetic properties respect to the $b$ axis, which can be due to a small anisotropy of the $g$-tensor for the V$^{4+}$ ions. A Curie-Weiss fit of the magnetic susceptibility above 130 K yields a positive Curie-Weiss temperature, indicating predominant small ferromagnetic correlations. At $T_{Cros}$ = 130±5 K the product $\chi(T)T$ shows the presence of a broad crossover, indicating a changing of exchange interactions character. The low-temperature magnetic behavior of $\beta$-TeVO$_4$ below 130 K clearly demonstrates the presence of predominant antiferromagnetic correlations in 1D spin system. In low-temperature behavior of magnetic susceptibility $\chi(T)$ the observed three features at $T$ = 2.28, 3.28, 4.65 K can be interpreted as a phase transitions of the studied quasi-one-dimensional spin system: to a long-range-ordered AF state at $T_N$ = 4.65 K and the further modifications of antiferromagnetic ordered phase (2.28; 3.28 K).

A fit of the magnetic susceptibility in frame of few low-dimensional spin-½ models was performed. The best qualitative agreement with the experiment in the wide temperature range 5 < $T$ < 400 K was obtained for an AF uniform exchange coupling model for spin-½ chain with the *single* intrachain antiferromagnetic coupling constant $J/k_B$ = 21.4±0.2 K. The using of the possible extensions of the 1D spin model shows that the strength of other exchange integrals (AF or FM nature) does not exceed a few Kelvin for this quasi-one-dimensional antiferromagnet. Although the knowledge of the magnetic data is not enough to definitely determine all model parameters, we believe that the compound $\beta$-TeVO$_4$ is a promising candidate to study the particularities of a weak-coupled $J_1$-$J_2$ model, especially from the side of a one-dimensional helimagnets. Further investigations, such as inelastic neutron scattering, would be helpful in determining the weak coupling parameters, magnetic ground state and the nature of the observed phase transformations. Given the lack of inorganic materials exhibiting this physics so far, the low temperature properties of this system are likely to attract a lot of attention in the future.

**ACKNOWLEDGMENTS**

This work was supported by the Ukrainian Fundamental Research State Fund (F4-11). We thank A.A. Zvyagin for important discussions.

REFERENCES
*Author to whom correspondence should be addressed: vpashchenko@ilt.kharkov.ua
[1] M. Isobe and Y. Ueda, J. Phys. Soc. Jpn. **65**, 1178 (1996); D.C. Johnston, J.W. Johnson,
D.P. Goshorn, A.J. Jacobson, Phys. Rev. B **35**, 219 (1987).
[2] M. Onoda and N. Nishiguchi, J. Phys. Condens. Matter **11**, 749 (1999).
[3] M. Onoda and N. Nishiguchi, J. Solid State Chem. **127**, 359 (1996).
[4] V. Gnezdilov, P. Lemmens, A.A. Zvyagin, V.O. Cheranovskii, K. Lamonova, Yu.G. Pashkevich, R.K. Kremer and H. Berger, Phys. Rev. B **78**, 184407 (2008).
[5] G. Meunier, J. Darriet, J. Galy, J. Sol. Stat. Chem. **6**, 67-73 (1973).
[6] S. Boudin, A. Guesdon, A. Leclaire, M.-M. Borel, Int. J. Inorg. Mater. **2**, 561 (2000).
[7] R.V. Panin, R.V. Shpanchenko, A.V. Mironov, Yu.A. Velikodny, E.V. Antipov, J. Hadermann, V.A. Tarnopolsky, A.B. Yaroslavtsev, E.E. Kaul and C. Geibel, Chem. Mater. **16**, 1048 (2004).
[8] M. Doublet and M. Lepetit, Phys. Rev. B **71**, 075119 (2005).
[9] Z.V. Popović, M.J. Konstantinović, R. Gajić, V.N. Popov, M. Isobe, Y. Ueda and
V.V. Moshchalkov, Phys. Rev. B **65**, 184303 (2002).
[10] J. Spitaler, E.Ya. Sherman, C. Ambrosch-Draxl, H.-G. Evertz, Phys. Scr. **T109**, 159 (2004).
[11] T. Masuda, A. Zheludev, B. Roessli, A. Bush, M. Markina and A. Vasiliev, Phys. Rev. B **72**, 014405 (2005).
[12] L. Capogna, M. Mayr, P. Horsch, M. Raichle, R. K. Kremer, M. Sofin, A. Maljuk, M. Jansen and B. Keimer, Phys. Rev. B **71**, 140402R (2005).
[13] R. Boča, *Current Methods in Inorganic Chemistry, Vol. I, Theoretical Foundations of Molecular Magnetism*, Elsevier, Amsterdam, Appendix 4, pp. 852-855 (1999) ; E. König, *Magnetic Properties of Coordination and Organo-Metallic Transition Metal Compounds*, in : Landolt-Börnstein, Neue Serie, Vol. II/2, Springer, Berlin, pp. 1-16 (1966).
[14] A. Abraham, B. Bleaney, Clarendon press, Oxford, 1970.
[15] D.C. Johnston, R.K. Kremer, M. Troyer, X. Wang, A. Klümper, S.L. Bud'ko, A.F. Panchula and P.C. Canfield, Phys. Rev. B **61**, 9558 (2000).

[16] J. Bonner, M. Fisher, Phys. Rev. **135**, A640 (1964).
[17] S. Zvyagin, G. Cao, Y. Xin, S. McCall, T. Caldwell, W. Moulton, L.-C. Brunel, A. Angerhofer, and J. E. Crow, Phys. Rev. B **66**, 064424 (2002).
[18] H.A. Bethe, Z. Phys. **71**, 205 (1931).
[19] S. Eggert, I. Affleck, and M. Takahashi, Phys. Rev. Lett. **73**, 332 (1994).
[20] H. J. Schulz, Phys. Rev. Lett. **77**, 2790 (1996).
[21] J. Borras-Almenar, E. Coronado, J. Curely, R. Georges, J.C. Gianduzzo, Inorg. Chem. **33**, 5171 (1994).
[22] B. Bleaney, K.D. Bowers , Proc. Roy.Soc. (London) Ser. A **214**, 451 (1952).